\documentclass[11pt,graphicx]{article}
\usepackage{amssymb,amsmath,amsfonts}
\usepackage{graphicx}
\usepackage{graphics}
\usepackage{eepic,epsfig}

=1.60cm

\begin{document}

\title{Vacuum polarization by a cosmic string in de Sitter spacetime }
\author{E. R. Bezerra de Mello$^{1}$\thanks{%
E-mail: emello@fisica.ufpb.br}\, and A. A. Saharian$^{2,3}$\thanks{%
E-mail: saharian@ictp.it} \\
\\
\textit{$^1$Departamento de F\'{\i}sica-CCEN, Universidade Federal da Para%
\'{\i}ba}\\
\textit{58.059-970, Caixa Postal 5.008, Jo\~{a}o Pessoa, PB, Brazil}\vspace{%
0.3cm}\\
\textit{$^2$Department of Physics, Yerevan State University,}\\
\textit{1 Alex Manoogian Street, 0025 Yerevan, Armenia}\vspace{0.3cm}\\
\textit{$^3$The Abdus Salam International Centre for Theoretical Physics,} \\
\textit{11 Strada Costiera, 34014 Trieste, Italy}}
\maketitle

\begin{abstract}
In this paper we investigate the vacuum polarization effect associated with
a quantum massive scalar field in a higher dimensional de Sitter spacetime
in the presence of a cosmic string. Because this investigation has been
developed in a pure de Sitter space, here we are mainly interested on the
effects induced by the presence of the string. So this analysis is developed
by expressing the complete Wightman function as the sum of two terms: The
first one corresponds to the bulk where the cosmic string is absent and the
second one is induced by the presence of the string. By using the Abel-Plana
summation formula, we show that for points away from the string the latter
is finite at the coincidence limit and it is used to evaluate the vacuum
averages of the square of the field and the energy-momentum tensor induced
by the cosmic string. Simple asymptotic formulae are obtained for these
expectation values for points near the string and at large distances from
it. It is shown that, depending on the curvature radius of de Sitter
spacetime, two regimes are realized with monotonic and oscillatory behavior
of the vacuum expectation values at large distances.
\end{abstract}

\bigskip


\bigskip

\section{Introduction}

De Sitter (dS) space is the curved spacetime which has been most studied in
quantum field theory during the past two decades. The main reason resides in
the fact that it is maximally symmetric and several physical problems can be
exactly solvable on this background; \footnote{%
De Sitter space enjoys the same degree of symmetry as the Minkowski one \cite%
{BD}} moreover the importance of these theoretical analysis increased by the
appearance of the inflationary cosmology scenario \cite{Linde}. In great
number of inflationary models, approximated dS spacetime is employed to
solve relevant problems in standard cosmology. During an inflationary epoch,
quantum fluctuations in the inflaton field introduce inhomogeneities and may
affect the transition toward the true vacuum. These fluctuations play
important role in the generation of cosmic structures from inflation. More
recently astronomical observations of high redshift supernovae, galaxy
clusters and cosmic microwave background \cite{Ries} indicate that at the
present epoch, the Universe is accelerating and can be well approximated by
a world with a positive cosmological constant.

Cosmic strings are linear topologically stable gravitational defects which
appear in the framework of grand unified theories. These objects could be
produced in very early Universe as a consequence of spontaneous breakdown of
gauge symmetry \cite{Kibble,V-S}. Although recent observations data on the
cosmic microwave background have ruled out cosmic strings as the primary
source for primordial density perturbation, they are still candidate for the
generation of a number of interesting physical effects such as gamma ray
burst \cite{Berezinski}, gravitational waves \cite{Damour} and high energy
cosmic rays \cite{Bhattacharjee}. Recently, cosmic strings have attracted
renewed interest partly because a variant of their formation mechanism is
proposed in the framework of brane inflation \cite{Sarangi}-\cite{Dvali}.

Many of high energy theories of fundamental physics are formulated in higher
dimensional spacetimes. Although topological defects have been first
analyzed in four-dimensional spacetime \cite{Kibble,V-S}, they have been
considered in the context of braneworld as well. In this scenario the
defects live in a $n$-dimensions submanifold embedded in a $(4+n)$%
-dimensional Universe. The cosmic string case corresponds to two additional
extra dimensions. In this context the gravitational effect of global strings
has been considered in \cite{Cohen,Ruth} as responsible for the
compactification from six to four spacetime dimensions, naturally producing
the observed hierarchy between electroweak and gravitational forces. The
analysis of quantum effects for various spin fields on the background of dS
spacetime, have been discussed by several authors (see, for instance, \cite%
{BD} and \cite{Cher68}-\cite{Mello} and references therein). Also for the
cosmic string spacetime these analysis have considered in \cite{scalar}-\cite%
{scalar5} and \cite{ferm}-\cite{Spinelly}, for scalar and fermionic fields
respectively.

In this paper, we shall investigate the one-loop quantum effects arising
from vacuum fluctuations associated with massive scalar field on the
background of higher dimensional de Sitter spacetime considering the
presence of a cosmic string in it (for cosmic strings in background of dS
spacetime see \cite{Line86}-\cite{Beze03}). Consequently these effects will
take into account the presence of the curvature of the spacetime and the
non-trivial topology associated with the two-dimensional conical sub-space.
The results obtained here can be used, in particular, for the investigation
of the effects of the quantum fluctuations induced by the string in the
inflationary phase. Though the cosmic strings produced in phase transitions
before or during early stages of inflation would have been drastically
diluted by the expansion, the formation of defects during inflation can be
triggered by a coupling of the symmetry breaking field to the inflaton field
or to the curvature of the background spacetime (see, for example, \cite{V-S}%
). The cosmic string can also be continuously created during inflation by
quantum-mechanical tunnelling \cite{Basu91}. Another class of models to
which the results of the present paper are applicable corresponds to
string-driven inflation where the cosmological expansion is driven entirely
by the string energy \cite{Turo88}.  The problem under consideration is also
of separate interest as an example with gravitational and topological
polarizations of the vacuum, where all calculations can be performed in a
closed form.

The paper is organized as follows. In section \ref{sec2} we present the
background associated with the geometry under consideration and the solution
of the Klein-Gordon equation admitting an arbitrary curvature coupling. Also
we present the complete Wightman function and consider the special case
where the planar angle deficit in the cosmic string subspace is an integer
fraction of $2\pi $. In sections \ref{sec3} and \ref{sec4} we calculate the
vacuum expectation values of the field squared and the energy-momentum
tensor induced by the cosmic string. Finally the main results of this paper
are summarized in section \ref{sec5}. Appendix contains some technical
details of the obtainment of the Wightaman function in a more workable form.

\section{Wightman function}

\label{sec2}

The main objective of this section is to calculate the positive frequency
Wightman function associated with a massive scalar field in a
higher-dimensional de Sitter spacetime taking into account the presence of a
cosmic string. This quantity is important to the calculation of scalar
vacuum averages. In order to do that we first obtain the complete set of
eigenfunctions for the Klein-Grodon equation admitting an arbitrary
curvature coupling. The geometry associated with the corresponding
background spacetime is given by the following line element:
\begin{equation}
ds^{2}=dt^{2}-e^{2t/\alpha }(dr^{2}+r^{2}d\phi
^{2}+\sum_{i=1}^{N}dz_{i}^{2})\ ,  \label{ds1}
\end{equation}%
where $r\geq 0$ and $\phi \in \lbrack 0,\ 2\pi /q]$ define the coordinates
on the conical geometry, $(t,\ z_{i})\in (-\infty ,\ \infty )$, and the
parameter $\alpha $ is related with the cosmological constant and Ricci
scalar by the formulae
\begin{equation}
\Lambda =\frac{(D-1)(D-2)}{2\alpha ^{2}},\;R=\frac{D(D-1)}{\alpha ^{2}}\ ,
\label{Lambda}
\end{equation}%
being $D$ the dimension of the spacetime given by $D=3+N$. The parameter $q$
is bigger than unity and codifies the presence of the cosmic string.
\footnote{%
It can be shown that by defining the coordinates $z_{0}=\alpha \sinh
(t/\alpha )+\frac{1}{2\alpha }e^{t/\alpha }(r^{2}+{\vec{x}}^{2})\ ,\
z_{D}=\alpha \cosh (t/\alpha )-\frac{1}{2\alpha }e^{t/\alpha }(r^{2}+{\vec{x}%
}^{2})\ ,z_{1}=e^{t/\alpha }r\cos \phi \ ,\ z_{2}=e^{t/\alpha }r\sin \phi \
,\ \mathrm{and}\ z_{i+2}=e^{t/\alpha }x_{i}$, they represent a hyperboloid $%
z_{0}^{2}-z_{1}^{2}-z_{2}^{2}-\sum z_{i}^{2}-z_{D}^{2}=-\alpha ^{2}$
embedded in a $(D+1)$-dimensional conical spacetime.} For further analysis,
in addition to the synchronous time coordinate $t$ we shall use the
conformal time $\tau $\ defined according to
\begin{equation}
\tau =-\alpha e^{-t/\alpha }\ ,\ -\infty <\ \tau \ <\ 0\ .  \label{tau}
\end{equation}%
In terms of this coordinate the above line element takes the form
\begin{equation}
ds^{2}=(\alpha /\tau )^{2}(d\tau ^{2}-dr^{2}-r^{2}d\phi
^{2}-\sum_{i=1}^{N}dz_{i}^{2})\ .  \label{ds2}
\end{equation}

The line element (\ref{ds1}) can also be written in static coordinates. For
simplicity we consider the corresponding coordinate transformation in the
case $D=4$ with $N=1$. This transformation is given by the relations
\begin{eqnarray}
t &=&t_{s}+\frac{\alpha }{2}\ln (1-r_{s}^{2}/\alpha ^{2}),\;r=\frac{%
r_{s}e^{-t_{s}/\alpha }\sin \theta }{\sqrt{1-r_{s}^{2}/\alpha ^{2}}},  \notag
\\
\;z_{1} &=&\frac{r_{s}e^{-t_{s}/\alpha }\cos \theta }{\sqrt{%
1-r_{s}^{2}/\alpha ^{2}}},\;\phi =\phi ,  \label{Coord}
\end{eqnarray}%
and the line element takes the form%
\begin{equation}
ds^{2}=(1-r_{s}^{2}/\alpha ^{2})dt_{s}^{2}-\frac{dr_{s}^{2}}{%
1-r_{s}^{2}/\alpha ^{2}}-r_{s}^{2}(d\theta ^{2}+\sin ^{2}\theta d\phi ^{2}),
\label{dSstatic}
\end{equation}%
with $0\leq \phi \leq 2\pi /q$. Introducing a new angular coordinate $%
\varphi =q\phi $, from (\ref{dSstatic}) we obtain the static line element of
the de Sitter spacetime with deficit angle previously discussed in \cite%
{Ghez02}. In this paper it is shown that to leading order in the
gravitational coupling the effect of the vortex on de Sitter spacetime is to
create a deficit angle in the metric (\ref{dSstatic}).

The field equation that will be considered is
\begin{equation}
(\nabla _{l}\nabla ^{l}+m^{2}+\xi R)\Phi (x)=0\ ,  \label{KG}
\end{equation}%
where $\xi $ is an arbitrary curvature coupling constant. The complete set
of solutions of this equation in the coordinate system defined by (\ref{ds2}%
) is:
\begin{equation}
\Phi _{\sigma }(x)=C_{\sigma }\eta ^{(D-1)/2}H_{\nu }^{(1)}(\lambda \eta
)J_{q|n|}(pr)e^{i\mathbf{k}\cdot \mathbf{z}+in\phi }\ ,\;\eta =\alpha
e^{-t/\alpha },  \label{sol1}
\end{equation}%
where $\lambda =\sqrt{p^{2}+k^{2}}$, $k=|\mathbf{k}|$,
\begin{equation}
\ \nu =\sqrt{(D-1)^{2}/4-\xi D(D-1)-m^{2}\alpha ^{2}},\ \ n=0,\pm 1,\pm 2,\
...  \label{nu}
\end{equation}%
In (\ref{sol1}), $H_{\nu }^{(1)}$ and $J_{\nu }$ represent the Hankel and
Bessel functions respectively, and $\sigma \equiv (p,\ \mathbf{k},\ n)$ is
the set of quantum numbers, being $p\in \lbrack 0,\ \infty )$. The
coefficient $C_{\sigma }$ can be found by the orthonormalization condition
\begin{equation}
-i\int d^{D-1}x\sqrt{|g|}g^{00}[\Phi _{\sigma }(x)\partial _{t}\Phi _{\sigma
^{\prime }}^{\ast }(x)-\Phi _{\sigma ^{\prime }}^{\ast }(x)\partial _{t}\Phi
_{\sigma }^{\ast }(x)]=\delta _{\sigma ,\sigma ^{\prime }}\ ,
\label{normcond}
\end{equation}%
where the integral is evaluated over the spatial hypersurface $\tau =\mathrm{%
const}$, and $\delta _{\sigma ,\sigma ^{\prime }}$ represents the
Kronecker-delta for discrete indices and Dirac-delta function for continuous
ones. This leads to the result
\begin{equation}
C_{\sigma }{}^{2}=\frac{qp\ e^{i(\nu -\nu ^{\ast })\pi /2}}{8\alpha
^{D-2}(2\pi )^{D-3}}\ ,  \label{coef}
\end{equation}%
for the normalization coefficient.

We shall employ the mode-sum formula to calculate the positive frequency
Wightman function:
\begin{equation}
G(x,x^{\prime })=\sum_{\sigma }\Phi _{\sigma }(x)\Phi _{\sigma }^{\ast
}(x^{\prime })\ .  \label{Green}
\end{equation}%
Substituting (\ref{sol1}), with respective coefficient (\ref{coef}), into (%
\ref{Green}) we obtain
\begin{eqnarray}
G(x,x^{\prime }) &=&\frac{q\ (\eta \eta ^{\prime })^{(D-1)/2}e^{i(\nu -\nu
^{\ast })\pi /2}}{8\alpha ^{D-2}(2\pi )^{D-3}}\int_{0}^{\infty }dp\ p\ \int d%
\mathbf{k}\,\,e^{i\mathbf{k}\cdot \Delta \mathbf{z}}  \notag \\
&&\times \sum_{n=-\infty }^{+\infty }e^{inq\Delta \phi
}J_{q|n|}(pr)J_{q|n|}(pr^{\prime })H_{\nu }^{(1)}(\lambda \eta )[H_{\nu
}^{(1)}(\lambda \eta ^{\prime })]^{\ast }\ ,  \label{Green1}
\end{eqnarray}%
with $\Delta \mathbf{z}=\mathbf{z}-\mathbf{z}^{\prime }$, $\Delta \phi =\phi
-\phi ^{\prime }$. In appendix \ref{sec:AppWF} we show that the expression
for the Wightman function can be transformed to the form%
\begin{equation}
G(x,x^{\prime })=\frac{q\ (\eta \eta ^{\prime })^{(D-1)/2}}{\pi
^{(D+1)/2}\alpha ^{D-2}}\ \sideset{}{'}{\sum}_{n=0}^{\infty }\cos (nq\Delta
\phi )\int_{0}^{\infty }du\,u^{(D-3)/2}e^{-\gamma u}I_{qn}(2rr^{\prime
}u)K_{\nu }(2\eta \eta ^{\prime }u)\ ,  \label{WF}
\end{equation}%
with
\begin{equation}
\gamma =|\Delta \mathbf{z}|^{2}+r^{2}+r^{\prime 2}-\eta ^{2}-\eta ^{\prime
2}\ .  \label{gamma}
\end{equation}%
In (\ref{WF}) the prime on the sign of summation means that the term $n=0$
should be halved.

The analysis of the vacuum polarization in the higher dimensional dS
spacetime, with toroidal compactification of extra dimensions, have been
developed in \cite{Saha-Set,Bell-Saha,Mello}. Here we are mainly interested
in quantum effects induced by the presence of the cosmic string. In order to
investigate these effects we introduce below the subtracted Wightman
function
\begin{equation}
G_{\mathrm{s}}(x,x^{\prime })=G(x,x^{\prime })-G(x,x^{\prime })|_{q=1}\ .
\label{gsub}
\end{equation}%
As the presence of the string does not change the curvature for the
background manifold for $r\neq 0$, the structure of the divergences in the
coincidence limit is the same for both terms on the right hand side. Hence,
for these points the function $G_{\mathrm{s}}(x,x^{\prime })$ is finite in
the coincidence limit. Regarding to the value of the parameter $q$, two
different approaches to obtain $G_{\mathrm{s}}(x,x^{\prime })$ will be
provided in the follows.

\subsection{Special case with an integer $q$}

The general expression for the scalar Wightman function, Eq. (\ref{WF}),
becomes a simpler one when $q$ is an integer number. In this case the
formula (\ref{WF}) is further simplified by using the relation~\cite{Pru}%
\begin{equation}
\sideset{}{'}{\sum}_{n=0}^{\infty }\cos (nq\Delta \phi )I_{qn}(2rr^{\prime
}u)=\frac{1}{2q}\sum_{k=0}^{q-1}\exp [2rr^{\prime }u\cos (\Delta \phi +2\pi
k/q)].  \label{Sumn1}
\end{equation}%
Now the integral over $u$ is evaluated with the help of formula%
\begin{eqnarray*}
\int_{0}^{\infty }du\,u^{(D-3)/2}e^{-cu}K_{\nu }(2\eta \eta ^{\prime }u) &=&%
\frac{2^{1-D}\sqrt{\pi }}{\Gamma (D/2)(\eta \eta ^{\prime })^{(D-1)/2}}%
\Gamma \left( \frac{D-1}{2}-\nu \right) \Gamma \left( \frac{D-1}{2}+\nu
\right) \\
&&\times F\left( \frac{D-1}{2}+\nu ,\frac{D-1}{2}-\nu ;\frac{D}{2};\frac{%
1-c/2\eta \eta ^{\prime }}{2}\right) ,
\end{eqnarray*}%
where $F(a,b;c;z)$ is the hypergeometric function.

Introducing the notation%
\begin{equation}
A_{D}=\frac{(4\pi )^{-D/2}}{\Gamma (D/2)}\Gamma \left( \frac{D-1}{2}-\nu
\right) \Gamma \left( \frac{D-1}{2}+\nu \right) ,  \label{AD}
\end{equation}%
the Wightman function is presented by a sum of $q$ images of the dS Wightman
function,
\begin{equation}
G(x,x^{\prime })=A_{D}\alpha ^{2-D}\sum_{k=0}^{q-1}\ G(c_{k}(x,x^{\prime
}))\ .  \label{Greenk}
\end{equation}%
In (\ref{Greenk}) we have introduced the notations%
\begin{equation}
G(z)=\ F\left( \frac{D-1}{2}+\nu ,\frac{D-1}{2}-\nu ;\frac{D}{2};1-z\right)
\ ,  \label{gz}
\end{equation}%
and
\begin{equation}
c_{k}(x,x^{\prime })=\frac{|\Delta \mathbf{z}|^{2}+r^{2}+r^{\prime
2}-2rr^{\prime }\cos (\Delta \phi +2\pi k/q)-(\Delta \eta )^{2}}{4\eta \eta
^{\prime }}.  \label{ck}
\end{equation}%
The $k=0$ term of (\ref{Greenk}) is divergent at the coincidence limit and
coincides with the Wightman function in dS spacetime. Finally, the part in
the Wightman function induced by the cosmic string, $G_{\mathrm{s}%
}(x,x^{\prime })$, is given by
\begin{equation}
G_{\mathrm{s}}(x,x^{\prime })=A_{D}\alpha ^{2-D}\sum_{k=1}^{q-1}\
G(c_{k}(x,x^{\prime }))\ ,  \label{Gsubk}
\end{equation}%
which is finite at the coincidence limit.

Note that by making use of the relation%
\begin{equation}
\ F\left( \frac{D-1}{2}+\nu ,\frac{D-1}{2}-\nu ;\frac{D}{2};\frac{1-u}{2}%
\right) =\frac{2^{D/2-1}\ \Gamma (D/2)}{(1-u^{2})^{(D-2)/4}}\ P_{\nu
-1/2}^{1-D/2}(u),  \label{FtoLeg}
\end{equation}%
we can express the function $G(z)$ in terms of the associated Legendre
function of the first kind. In the case of massless field and conformal
coupling,
\begin{equation}
\xi =\xi _{c}=\frac{D-2}{4(D-1)},  \label{conf}
\end{equation}%
one has $\nu =1/2$ and for the function (\ref{gz}) we find $G(z)=z^{1-D/2}$.

\subsection{General case}

For the general case of $q$ the subtracted Wightman function can be
expressed by%
\begin{eqnarray}
G_{\mathrm{s}}(x,x^{\prime }) &=&\frac{(\eta \eta ^{\prime })^{(D-1)/2}}{\pi
^{(D+1)/2}\alpha ^{D-2}}\ \int_{0}^{\infty }du\,u^{(D-3)/2}e^{-\gamma
u}K_{\nu }(2\eta \eta ^{\prime }u)  \notag \\
&&\times \sideset{}{'}{\sum}_{n=0}^{\infty }\left[ q\cos (nq\Delta \phi
)I_{qn}(2rr^{\prime }u)-\cos (n\Delta \phi )I_{n}(2rr^{\prime }u)\right] .
\label{WF2}
\end{eqnarray}%
A more convenient expression for this function can be provided by using the
Abel-Plana formula (see, for instance, \cite{SahaRev}) for the summation
over $n$:
\begin{equation}
\sideset{}{'}{\sum}_{n=0}^{\infty }F(n)=\int_{0}^{\infty }du\
F(u)+i\int_{0}^{\infty }du\ \frac{F(iu)-F(-iu)}{e^{2\pi u}-1}\ .
\label{AbelPlana}
\end{equation}%
Now we can see that in the evaluation of the difference the terms coming
from the first integral on the right of the Abel-Plana formula cancel out
and one obtains%
\begin{equation}
\sideset{}{'}{\sum}_{n=0}^{\infty }\left[ q\cos (nq\Delta \phi
)I_{qn}(z)-\cos (n\Delta \phi )I_{n}(z)\right] =\frac{2}{\pi }%
\int_{0}^{\infty }dv\,\cosh (v\Delta \phi )g(q,v)K_{iv}(z),  \label{sumn}
\end{equation}%
with the notation%
\begin{equation}
g(q,v)=\sinh (\pi v)\left( \frac{1}{e^{2\pi v/q}-1}-\frac{1}{e^{2\pi v}-1}%
\right) .  \label{gqv}
\end{equation}%
As a result the subtracted Wightman function takes the form%
\begin{eqnarray}
G_{\mathrm{s}}(x,x^{\prime }) &=&\frac{2(\eta \eta ^{\prime })^{\frac{D-1}{2}%
}}{\pi ^{\frac{D+3}{2}}\alpha ^{D-2}}\ \int_{0}^{\infty }du\,u^{\frac{D-3}{2}%
}e^{-\gamma u}K_{\nu }(2\eta \eta ^{\prime }u)  \notag \\
&&\times \int_{0}^{\infty }dv\,\cosh (v\Delta \phi )g(q,v)K_{iv}(2rr^{\prime
}u).  \label{WFsub}
\end{eqnarray}

For a conformally coupled massless scalar field we have $\nu =1/2$:%
\begin{equation}
G_{\mathrm{s}}(x,x^{\prime })=\frac{(\eta \eta ^{\prime })^{\frac{D-2}{2}}}{%
\pi ^{\frac{D+2}{2}}\alpha ^{D-2}}\ \int_{0}^{\infty }dv\,\cosh (v\Delta
\phi )g(q,v)\int_{0}^{\infty }du\,u^{\frac{D-4}{2}}e^{-(\gamma +2\eta \eta
^{\prime })u}K_{iv}(2rr^{\prime }u).  \label{WFsub1}
\end{equation}%
In this case the problem under consideration is conformally related to the
problem with a cosmic string in the Minkowski spacetime and the
corresponding subtracted Wightman functions are connected by the standard
conformal transformation.

\section{The computation of $\langle \Phi ^{2}\rangle $}

\label{sec3}

The formal expression for the vacuum expectation value (VEV) of the field
squared is given by evaluating the Wightman function at the coincidence
limit:
\begin{equation}
\langle \Phi ^{2}(x)\rangle =\lim_{x^{\prime }\rightarrow x}G(x,x^{\prime
})\ .  \label{phi2}
\end{equation}%
However this procedure provides a divergent result and to obtain a finite
and well defined value some renormalization procedure is necessary. The
important point here is that the presence of the cosmic string does not
introduce additional curvature. Thus, the divergences are contained in the
part corresponding to the pure dS spacetime and the string induced part is
finite for points outside the string. As we have already extracted the first
part, the renormalization procedure is reduced to the renormalization of the
dS part in the absence of the string which is already done in literature
\cite{Cand75,Dowk76,Bunc78}. So according to the fact that
\begin{equation}
G(x,x^{\prime })=G_{\mathrm{dS}}(x,x^{\prime })+G_{\mathrm{s}}(x,x^{\prime })
\label{WFdec}
\end{equation}%
we can conclude that
\begin{equation}
\langle \Phi ^{2}\rangle =\langle \Phi ^{2}\rangle _{\mathrm{dS}}+\langle
\Phi ^{2}\rangle _{\mathrm{s}},\;\langle \Phi ^{2}\rangle _{\mathrm{s}}=G_{%
\mathrm{s}}(x,x)\ .  \label{phi2s}
\end{equation}%
Due to the maximal symmetry of the dS spacetime the VEV $\langle \Phi
^{2}\rangle _{\mathrm{dS}}$ does not depend on the spacetime point. In the
special case of $q$ being an integer number the resulting formulae are
essentially simplified and in the following we present the corresponding
calculations separately.

\subsection{Special case}

For integer $q$ we use (\ref{Gsubk}). The VEV is promptly obtained:%
\begin{equation}
\langle \Phi ^{2}\rangle _{\mathrm{s}}=A_{D}\alpha ^{2-D}\sum_{k=1}^{q-1}\
G(c_{k}),  \label{Phi}
\end{equation}%
with
\begin{equation}
c_{k}=y\sin ^{2}(\pi k/q),\;y\equiv r^{2}/\eta ^{2}.  \label{uk}
\end{equation}%
Note that the VEV (\ref{Phi}) is a function of the ratio $r/\eta $ which is
the proper distance from the string, $\alpha r/\eta $, measured in the units
of the dS curvature radius $\alpha $. In the discussion below, for the
evaluation of the VEV of the energy-momentum tensor we will need also the
covariant d'Alembertian for the string induced part in the VEV\ of the field
squared:%
\begin{equation}
\nabla _{\sigma }\nabla ^{\sigma }\langle \Phi ^{2}\rangle _{\mathrm{s}%
}=2A_{D}\alpha ^{-D}\sum_{k=1}^{q-1}c_{k}\left[ 2(1-1/y)[c_{k}G^{\prime
}(c_{k})]^{\prime }+(D-1)G^{\prime }(c_{k})\right] ,  \label{Dalambphi2}
\end{equation}%
where the prime means the derivative with respect to $c_{k}$. The second
derivative in this expression can be excluded by using the equation for the
function $G(z)$ which is easily obtained from the hypergeometric equation:%
\begin{equation}
(1-z)[zG^{\prime }(z)]^{\prime }-[1-D/2+(D-1)z]G^{\prime }(z)-\left[ \xi
D(D-1)+m^{2}\alpha ^{2}\right] G(z)=0.  \label{Geq}
\end{equation}%
For the first derivative of this function one has
\begin{equation}
G^{\prime }(z)=-\frac{2}{D}\left[ \xi D(D-1)+m^{2}\alpha ^{2}\right] F\left(
\frac{D+1}{2}+\nu ,\frac{D+1}{2}-\nu ;\frac{D}{2}+1;1-z\right) .
\label{Gder}
\end{equation}

\begin{figure}[tbph]
\begin{center}
\begin{tabular}{cc}
\epsfig{figure=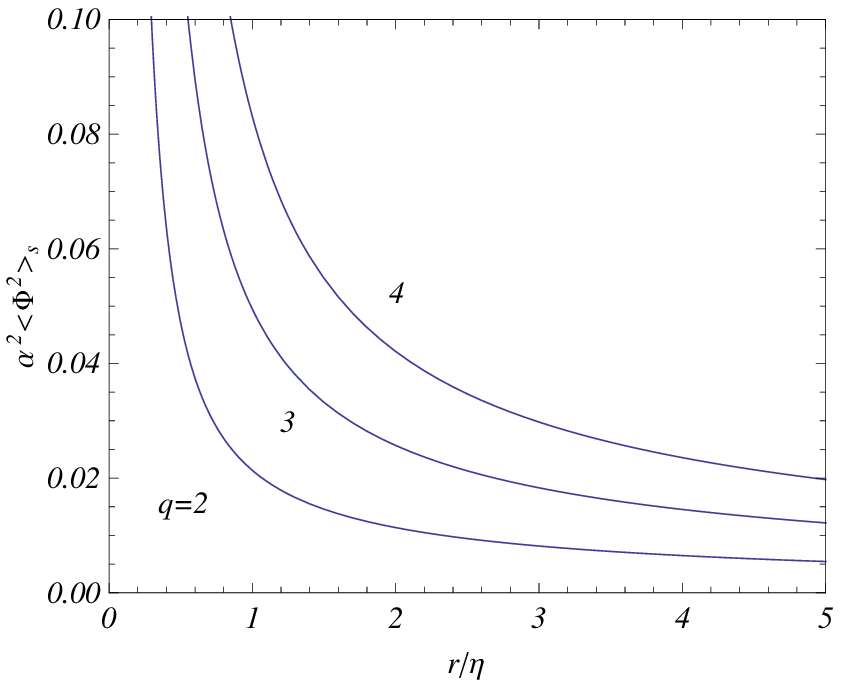,width=7.cm,height=6.cm} & \quad %
\epsfig{figure=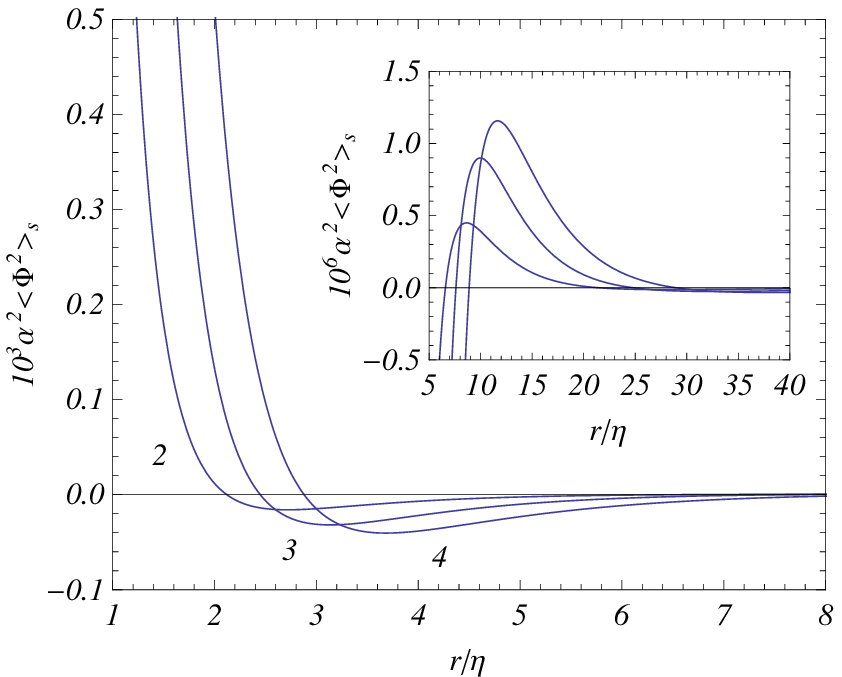,width=7.cm,height=6cm}%
\end{tabular}%
\end{center}
\caption{The string induced part in the VEV of the field squared as a
function of the ratio $r/\protect\eta $ for various values of the parameter $%
q$ for a minimally coupled scalar field with $m\protect\alpha =1$ (left
panel) and $m\protect\alpha =2$ (right panel) in 4-dimensional dS spacetime.}
\label{fig1}
\end{figure}

In figure \ref{fig1} we have plotted the string induced part in the VEV of
the field squared versus the ratio $r/\eta $ (proper distance from the
string measured in units of $\alpha $) for minimally coupled scalar field ($%
\xi =0$) with $m\alpha =1$ (left panel) and $m\alpha =2$ (right panel) in
4-dimensional spacetime. The numbers near the curves correspond to the
values of the parameter $q$. For the left panel the parameter $\nu $ is real
and for the right one this parameter is imaginary. In the latter case the
oscillatory behavior of the VEV is seen at large distances from the string.
A simple formula for the asymptotic behavior of the VEV of the field squared
at large distances from the string will be given below for the general case
of the parameter $q$.

For a conformally coupled massless scalar field form (\ref{Phi}) one finds%
\begin{equation}
\langle \Phi ^{2}\rangle _{\mathrm{s}}=\left( \frac{\eta }{\alpha r}\right)
^{D-2}\frac{\Gamma (D/2-1)}{(4\pi )^{D/2}}\sum_{k=1}^{q-1}\sin ^{2-D}(\pi
k/q)\ .  \label{phi2sconf}
\end{equation}%
For even values of $D$, the summation on the right-hand side of the above
expression can be obtained in a closed form by use of the formulae
\begin{equation}
I_{N+2}(x)=\frac{I_{N}^{\prime \prime }(x)+N^{2}I_{N}(x)}{N(N+1)},\ I_{2}(x)=%
\frac{q^{2}}{\sin ^{2}(qx)}-\frac{1}{\sin ^{2}(x)}\ ,  \label{IN}
\end{equation}%
for the sum
\begin{equation}
I_{N}(x)=\sum_{k=1}^{q-1}\sin ^{-N}(x+k\pi /q)\ .  \label{IN1}
\end{equation}%
In particular, for a 4-dimensional spacetime, $D=4$, we need $%
I_{2}(0)=(q^{2}-1)/3$. Consequently,
\begin{equation}
\langle \Phi ^{2}\rangle _{\mathrm{s}}=\frac{q^{2}-1}{48\pi ^{2}}\left(
\frac{\eta }{\alpha r}\right) ^{2}\ .  \label{D.1}
\end{equation}%
For a six-dimensional spacetime, $I_{4}(x)$ is a long expression obtained by
the recurrence relation above, however we find $%
I_{4}(0)=(q^{2}-1)(q^{2}+11)/45$. In this case we have
\begin{equation}
\langle \Phi ^{2}\rangle _{\mathrm{s}}=\frac{(q^{2}-1)(q^{2}+11)}{2880\pi
^{3}}\left( \frac{\eta }{\alpha r}\right) ^{4}\ .  \label{D.2}
\end{equation}%
The above results, Eqs. (\ref{D.1}) and (\ref{D.2}), are analytical
functions of $q$, and by the analytical continuation they are valid for
arbitrary values of $q$.

\subsection{General case}

For the general case of the parameter $q$ we have an integral representation
for the VEV of the field squared by using (\ref{WFsub}):%
\begin{equation}
\langle \Phi ^{2}\rangle _{\mathrm{s}}=\frac{8\alpha ^{2-D}}{(2\pi
)^{(D+3)/2}}\int_{0}^{\infty }dv\,g(q,v)\ \int_{0}^{\infty }dz\frac{e^{-z}}{z%
}K_{iv}(z)F(z\eta ^{2}/r^{2}),  \label{phi2gen}
\end{equation}%
\newline
where and in the discussion below we use the notation%
\begin{equation}
F(x)=x^{\frac{D-1}{2}}e^{x}K_{\nu }(x).  \label{F(u)}
\end{equation}%
As before, we see that the contribution induced by the cosmic string in the
VEV of the field squared depends on the ratio $r/\eta $. For the covariant
d'Alembertian of the string induced part in the VEV of the field squared we
find%
\begin{eqnarray}
\nabla _{\sigma }\nabla ^{\sigma }\langle \Phi ^{2}\rangle _{\mathrm{s}} &=&%
\frac{16\alpha ^{-D}}{(2\pi )^{(D+3)/2}}\int_{0}^{\infty }dv\,g(q,v)\
\int_{0}^{\infty }dz\,e^{-yz}K_{iv}(yz)  \notag \\
&&\times \left\{ 2(1-1/y)[zF^{\prime }(z)]^{\prime }+(1-D)F^{\prime
}(z)\right\} .  \label{Dalphi2Gen}
\end{eqnarray}%
Note that for the derivatives of the function $F(z)$ we have the relations
\begin{equation}
\lbrack zF^{\prime }(z)]^{\prime }=\left( D-1+2z\right) F^{\prime }(z)-\left[
\xi D(D-1)+m^{2}\alpha ^{2}+(D-2)z\right] F(z)/z,  \label{Fder}
\end{equation}%
and%
\begin{equation}
F^{\prime }(z)=z^{\frac{D-1}{2}}e^{z}\left[ \left( \frac{D-1}{2z}+\frac{\nu
}{z}+1\right) K_{\nu }(z)-K_{\nu +1}(z)\right] .  \label{Fder1}
\end{equation}

For a conformally coupled massless scalar field, $\nu =1/2$, the formula (%
\ref{phi2gen}) reduces to
\begin{equation}
\langle \Phi ^{2}\rangle _{\mathrm{s}}=\frac{8(4\pi )^{-(D+1)/2}}{\Gamma
((D-1)/2)}\left( \frac{\eta }{\alpha r}\right) ^{D-2}\int_{0}^{\infty
}dv\,g(q,v)\ \Gamma (D/2-1+iv)\Gamma (D/2-1-iv).  \label{phi2conf}
\end{equation}%
For even values of $D$, by using the formula $\Gamma (1+iv)\Gamma (1-iv)=\pi
v/\sinh (\pi v)$, the integral is evaluated explicitly. In particular, for $%
D=4,6$ we recover the results (\ref{D.1}) and (\ref{D.2}).

The general formula for the string induced part in the VEV of the field
squared is simplified in asymptotic regions of small and large distances.
For small values of the ratio $r/\eta $ (points near the cosmic string) the
argument of the Mac-Donald function in (\ref{phi2gen}) is large and to the
leading order one has $F(x)\approx \sqrt{\pi /2}x^{D/2-1}$. With this
approximation the integral over $z$ is evaluated by using the formula \cite%
{Pru}
\begin{equation}
\int_{0}^{\infty }dz\,z^{\beta -1}e^{-z}K_{iv}(z)=\frac{\sqrt{\pi }}{%
2^{\beta }}\frac{\Gamma (\beta +iv)\Gamma (\beta -iv)}{\Gamma (\beta +1/2)},
\label{Intform2}
\end{equation}%
and one finds%
\begin{equation}
\langle \Phi ^{2}\rangle _{\mathrm{s}}\approx \left( \frac{\eta }{\alpha r}%
\right) ^{D-2}\int_{0}^{\infty }dv\,g(q,v)h(v).  \label{phi2small}
\end{equation}%
In this formula and in the discussion below we use the notation%
\begin{equation}
h(v)=\frac{8(4\pi )^{-(D+1)/2}}{\Gamma ((D-1)/2)}\Gamma (D/2-1+iv)\Gamma
(D/2-1-iv).  \label{hv}
\end{equation}%
Comparing with (\ref{phi2conf}), we conclude that at small distance, the
leading order term of the string induced part in the VEV of the field
squared coincides with the corresponding result for a massless conformally
coupled field.

At large distances from the string, $r/\eta \gg 1$, we replace the
Mac-Donald function in $F(x)$ by the corresponding asymptotic expression for
small values of the argument. After that the integral over $z$ is evaluated
by formula (\ref{Intform2}) and for real values of the parameter $\nu $ one
finds%
\begin{eqnarray}
\langle \Phi ^{2}\rangle _{\mathrm{s}} &\approx &\frac{\alpha ^{2-D}\Gamma
(\nu )}{\pi ^{D/2+1}\Gamma (D/2-\nu )}\left( \frac{\eta }{2r}\right)
^{D-1-2\nu }\int_{0}^{\infty }dv\,g(q,v)  \notag \\
&&\times \Gamma (\frac{D-1}{2}-\nu +iv)\Gamma (\frac{D-1}{2}-\nu -iv).
\label{phi2large}
\end{eqnarray}%
For a conformally coupled massless scalar field this result coincides with
the exact formula (\ref{phi2conf}). For imaginary values of $\nu $, in the
similar way, we have the following asymptotic estimate%
\begin{equation}
\langle \Phi ^{2}\rangle _{\mathrm{s}}\approx -\frac{4B(q,|\nu |)}{(4\pi
)^{D/2}\alpha ^{D-2}}\left( \frac{\eta }{r}\right) ^{D-1}\sin [2|\nu |\ln
(\eta /2r)+\psi _{0}],  \label{phi2largeIm}
\end{equation}%
where $B(q,|\nu |)$ and the phase $\psi _{0}$ are defined by the relation%
\begin{equation}
B(q,|\nu |)e^{i\psi _{0}}=\int_{0}^{\infty }dv\,g(q,v)\ \frac{\Gamma
((D-1)/2+\nu +iv)\Gamma ((D-1)/2+\nu -iv)}{\sinh (|\nu |\pi )\Gamma (1+\nu
)\Gamma (D/2+\nu )}.  \label{Bpsi0}
\end{equation}%
Hence, in this case, at large distances from the string, the string induced
part decays as $r^{1-D}$ and the damping of the corresponding VEV has an
oscillatory nature.

\section{Vacuum expectation value of the energy-momentum tensor}

\label{sec4}

In this section we shall analyze the contribution to the VEV of the scalar
energy-momentum tensor induced by the cosmic string. As in the previous
section we may write
\begin{equation}
\langle T_{\mu \nu }\rangle =\langle T_{\mu \nu }\rangle _{\mathrm{dS}%
}+\langle T_{\mu \nu }\rangle _{\mathrm{s}}\ ,  \label{Tmudec}
\end{equation}%
where $\langle T_{\mu \nu }\rangle _{\mathrm{dS}}$ is the corresponding VEV
in dS spacetime when the string is absent. The latter does not depend on the
spacetime point and is well-investigated in literature \cite%
{Cand75,Dowk76,Bunc78}. It corresponds to a gravitational source of the
cosmological constant type and, in combination with the initial cosmological
constant $\Lambda $, given by (\ref{Lambda}), leads to the effective
cosmological constant $\Lambda _{\mathrm{eff}}=\Lambda +8\pi G\langle
T_{0}^{0}\rangle _{\mathrm{dS}}$, where $G$\ is the Newton gravitational
constant. The string induced contribution is obtained by use of the formula
\begin{equation}
\langle T_{\mu \nu }\rangle _{\mathrm{s}}=\lim_{x^{\prime }\rightarrow
x}\partial _{\mu ^{\prime }}\partial _{\nu }G_{\mathrm{s}}(x,x^{\prime })+%
\left[ \left( \xi -1/4\right) g_{\mu \nu }\nabla _{\sigma }\nabla ^{\sigma
}-\xi \nabla _{\mu }\nabla _{\nu }-\xi R_{\mu \nu }\right] \langle \Phi
^{2}\rangle _{\mathrm{s}}\ ,  \label{EMT}
\end{equation}%
where
\begin{equation}
R_{\mu \nu }=(D-1)g_{\mu \nu }/\alpha ^{2}  \label{Rmu}
\end{equation}%
is the Ricci tensor for the background spacetime.

\subsection{Special case}

First we consider the special case of integer $q$. By using the expressions
for the Wightman function and the VEV of the field squared, after long
calculations the diagonal components of the energy-momentum tensor are
presented in the form (no summation over $\mu $)%
\begin{equation}
\langle T_{\mu }^{\mu }\rangle _{\mathrm{s}}=\frac{A_{D}}{\alpha ^{D}}%
\sum_{k=1}^{q-1}f^{(\mu )}(c_{k})+A,  \label{Tmusp}
\end{equation}%
where we have defined
\begin{equation}
A=\left[ \left( \xi -1/4\right) \nabla _{\sigma }\nabla ^{\sigma }-\xi
(D-1)/\alpha ^{2}\right] \langle \Phi ^{2}\rangle _{\mathrm{s}}.
\label{Aemt}
\end{equation}%
In (\ref{Tmusp}) we have introduced the notations%
\begin{eqnarray}
f^{(0)}(z) &=&(1-4\xi )z[zG^{\prime }(z)]^{\prime }+G^{\prime }(z)/2,  \notag
\\
f^{(1)}(z) &=&(4\xi -1)z\left[ zG^{\prime }(z)\right] ^{\prime }/y+\left[
1-4\xi z(1/y+1)\right] G^{\prime }(z)/2,  \label{f1} \\
f^{(2)}(z) &=&(1-z/y)\left[ zG^{\prime }(z)\right] ^{\prime }-\left[ 1+4\xi
z(1-1/y)\right] G^{\prime }(z)/2,  \notag \\
f^{(\mu )}(z) &=&(1-4\xi z)G^{\prime }(z)/2,\;\mu =3,\ldots ,D-1.  \notag
\end{eqnarray}

For the non-zero off-diagonal component one has%
\begin{equation}
\langle T_{1}^{0}\rangle _{\mathrm{s}}=\frac{\eta A_{D}}{r\alpha ^{D}}%
\sum_{k=1}^{q-1}c_{k}\left\{ (1-4\xi )[c_{k}G^{\prime }(c_{k})]^{\prime }\
+2\xi G^{\prime }(c_{k})\right\} .  \label{T01sp}
\end{equation}%
We recall that for a conformally coupled massless scalar field one has $%
G(z)=z^{1-D/2}$ and, as it can be easily seen from (\ref{T01sp}), the
off-diagonal component vanishes. The explicit expression for the part $A$
directly follows from formulae (\ref{Phi}) and (\ref{Dalambphi2}):%
\begin{equation}
A=\frac{A_{D}}{\alpha ^{D}}\sum_{k=1}^{q-1}\left\{ \left( 4\xi -1\right)
c_{k}\left[ \left( 1-\frac{1}{y}\right) [c_{k}G^{\prime }(c_{k})]^{\prime }+%
\frac{D-1}{2}G^{\prime }(c_{k})\right] -\xi (D-1)\ G(c_{k})\right\} .
\label{Aspecial}
\end{equation}

Now it can be checked that the components of the energy-momentum tensor
satisfy the trace relation
\begin{equation}
\langle T_{\mu }^{\mu }\rangle _{\mathrm{s}}=D(\xi -\xi _{c})\nabla _{\mu
}\nabla ^{\mu }\langle \Phi ^{2}\rangle _{\mathrm{s}}+m^{2}\langle \Phi
^{2}\rangle _{\mathrm{s}}.  \label{trace}
\end{equation}%
In particular, for a conformally coupled massless scalar this tensor is
traceless and the trace anomalies are contained in the purely dS part only.
Note that from the covariant conservation equation $\nabla _{\nu }T_{\mu
}^{\nu }=0$ we have also the following relations between the components of
the energy-momentum tensor:%
\begin{eqnarray}
\partial _{\eta }T_{0}^{0}+\frac{1}{\eta }\left( T_{\mu }^{\mu
}-DT_{0}^{0}\right) -\frac{1}{r}\partial _{r}(T_{0}^{1}r) &=&0,  \notag \\
\partial _{\eta }T_{1}^{0}+\frac{D}{\eta }T_{1}^{0}-\frac{1}{r}\partial
_{r}(T_{1}^{1}r)+\frac{1}{r}T_{2}^{2} &=&0.  \label{conteq}
\end{eqnarray}%
These relations are fullfilled separately for pure and string induced parts
in the VEV of the energy-momentum tensor.

\begin{figure}[tbph]
\begin{center}
\begin{tabular}{cc}
\epsfig{figure=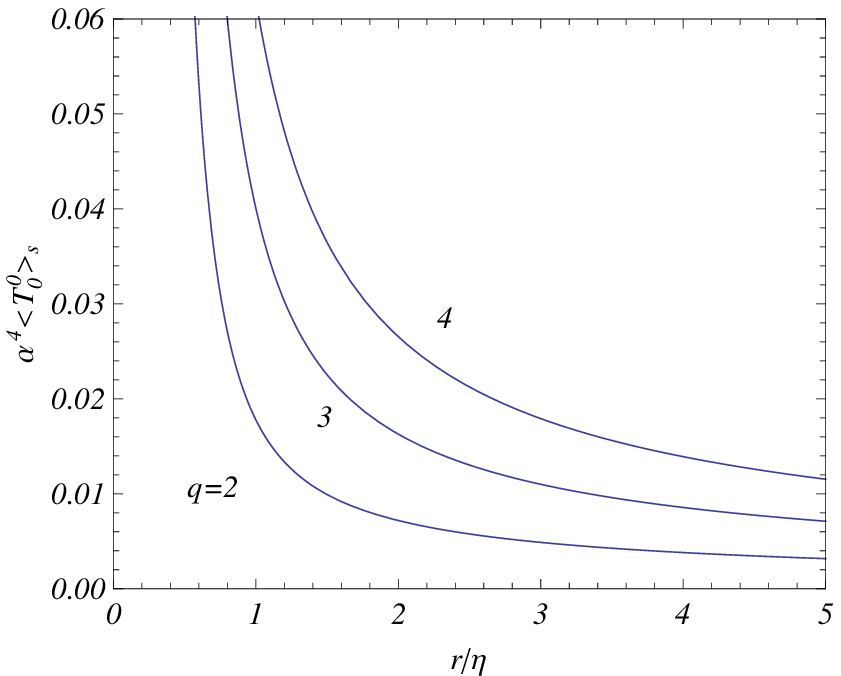,width=7.cm,height=6.cm} & \quad %
\epsfig{figure=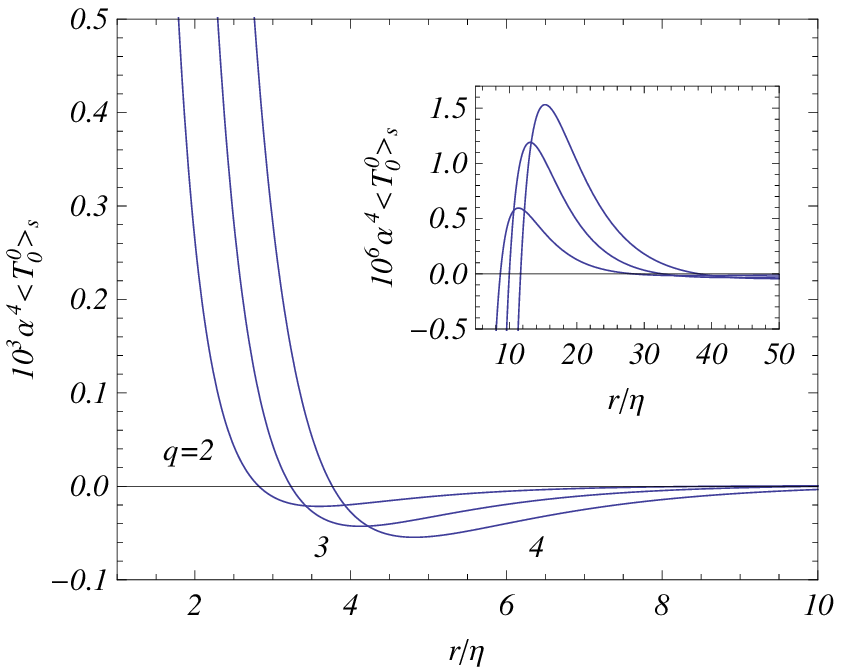,width=7.cm,height=6cm}%
\end{tabular}%
\end{center}
\caption{The vacuum energy density induced by the string in 4-dimensional dS
spacetime versus $r/\protect\eta $ for a minimally coupled scalar field with
$m\protect\alpha =1$ (left panel) and $m\protect\alpha =2$ (right panel).}
\label{fig2}
\end{figure}

In figure \ref{fig2} the string induced part in the VEV of the energy
density is depicted as a function of $r/\eta $ in the case of a minimally
coupled scalar field with $m\alpha =1$ (left panel) and $m\alpha =2$ (right
panel) in 4-dimensional dS spacetime. For the right panel the parameter $\nu
$ is imaginary and the decay of the VEV is oscillatory. In the next
subsection this feature will be seen from the corresponding asymptotic
formula.

\subsection{General case}

\label{4.2}

For an arbitrary value of $q$, we need the formulae (\ref{WFsub}) and (\ref%
{phi2gen}). For this case the calculations of the components of the
energy-momentum tensor are much longer. The final results for the diagonal
components are given below (no summation over $\mu $):%
\begin{equation}
\langle T_{\mu }^{\mu }\rangle _{\mathrm{s}}=A-\frac{8\alpha ^{-D}}{(2\pi
)^{(D+3)/2}}\ \int_{0}^{\infty }dv\,g(q,v)\int_{0}^{\infty
}dz\,e^{-yz}K_{iv}(yz)g^{(\mu )}(z),  \label{Tmu}
\end{equation}%
where $A$ is defined by the relation (\ref{Aemt}) and the following
notations are introduced%
\begin{eqnarray}
g^{(0)}(z) &=&(4\xi -1)[zF^{\prime }(z)]^{\prime }+F(z),  \notag \\
g^{(1)}(z) &=&(1-4\xi )[zF^{\prime }(z)]^{\prime }/y-2\xi (1/y+1)F^{\prime
}(z)+F(z),  \label{g1} \\
g^{(2)}(z) &=&[zF^{\prime }(z)]^{\prime }/y-2\,\left[ \,z+\xi (1-1/y)\right]
F^{\prime }(z)-F(z),  \notag \\
g^{(\mu )}(z) &=&-2\xi F^{\prime }(z)+F(z),\;\mu =3,\ldots ,D-1.  \notag
\end{eqnarray}%
In addition to diagonal components there is also non-zero off-diagonal
component%
\begin{equation}
\langle T_{1}^{0}\rangle _{\mathrm{s}}=-\frac{8\alpha ^{-D}\eta /r}{(2\pi
)^{(D+3)/2}}\ \int_{0}^{\infty }dv\,g(q,v)\int_{0}^{\infty
}dz\,K_{iv}(zy)e^{-zy}\left\{ (4\xi -1)[zF^{\prime }(z)]^{\prime }+2\xi
F^{\prime }(z)\right\} .  \label{T01}
\end{equation}%
Note that for the covariant d'Alembertian in (\ref{Aemt}) we have formula (%
\ref{Dalphi2Gen}) and for this part one obtains%
\begin{eqnarray}
A &=&\frac{8\alpha ^{-D}}{(2\pi )^{(D+3)/2}}\int_{0}^{\infty }dv\,g(q,v)\
\int_{0}^{\infty }dze^{-yz}K_{iv}(yz)  \notag \\
&&\times \left\{ \left( 4\xi -1\right) \left[ \left( 1-\frac{1}{y}\right)
[zF^{\prime }(z)]^{\prime }+\frac{1-D}{2}F^{\prime }(z)\right] -\xi \frac{D-1%
}{z}F(z)\right\} .  \label{Agen}
\end{eqnarray}
It can be explicitly checked that the string induced parts in the components
of the VEV of the energy-momentum tensor satisfy the trace relation (\ref%
{trace}).

For a conformally coupled massless scalar field one has $\nu =1/2$ and the
off-diagonal component (\ref{T01}) vanishes. For the diagonal components in
this case one has (no summation over $\mu $)
\begin{equation}
\langle T_{\mu }^{\mu }\rangle _{\mathrm{s}}=\frac{\langle T_{2}^{2}\rangle
_{\mathrm{s}}}{1-D}=-\frac{1}{D-1}\left( \frac{\eta }{\alpha r}\right)
^{D}\int_{0}^{\infty }dv\,v^{2}g(q,v)h(v).  \label{Tconf}
\end{equation}%
where the function $h(v)$ is defined by relation (\ref{hv}).

The asymptotic behavior of the string induced part in the VEV of the
energy-momentum tensor at small and large distances is investigated in the
way similar to that used above for the case of the field squared. For points
near the string, $r/\eta \ll 1$, to the leading order we have%
\begin{eqnarray}
\langle T_{0}^{0}\rangle _{\mathrm{s}} &\approx &\langle T_{3}^{3}\rangle _{%
\mathrm{s}}\approx -\left( \frac{\eta }{\alpha r}\right)
^{D}\int_{0}^{\infty }dv\,g(q,v)h(v)\left[ (D-2)^{2}\left( \xi -\xi
_{c}\right) +\frac{v^{2}}{D-1}\right] ,  \label{T00small} \\
\langle T_{1}^{1}\rangle _{\mathrm{s}} &\approx &\frac{\langle
T_{2}^{2}\rangle _{\mathrm{s}}}{1-D}\approx \left( \frac{\eta }{\alpha r}%
\right) ^{D}\int_{0}^{\infty }dv\,g(q,v)h(v)\left[ (D-2)\left( \xi -\xi
_{c}\right) -\frac{v^{2}}{D-1}\right] ,  \label{T11small}
\end{eqnarray}%
For the off-diagonal component the corresponding formula takes the form%
\begin{equation}
\langle T_{1}^{0}\rangle _{\mathrm{s}}\approx -(D-2)\left( \xi -\xi
_{c}\right) \frac{D-1}{\alpha }\left( \frac{\eta }{\alpha r}\right) ^{D-1}\
\int_{0}^{\infty }dv\,g(q,v)h(v).  \label{T01small}
\end{equation}%
In the case of a conformally coupled massless scalar field these formulae
coincide with the exact results.

At large distances form the string, $r/\eta \gg 1$, and for real values of
the parameter $\nu $\ for the energy density we have the following
asymptotic expression:%
\begin{eqnarray}
\langle T_{0}^{0}\rangle _{\mathrm{s}} &\approx &-\left[ \left( D-1-2\nu
\right) \left( \xi -1/4\right) +\xi \right] \frac{(D-1)\alpha ^{-D}\Gamma
(\nu )}{\pi ^{D/2+1}\Gamma (D/2-\nu )}\left( \frac{\eta }{2r}\right)
^{D-1-2\nu }\   \notag \\
&&\times \int_{0}^{\infty }dv\,g(q,v)\Gamma \left( \frac{D-1}{2}-\nu
+iv\right) \Gamma \left( \frac{D-1}{2}-\nu -iv\right) .  \label{T00large}
\end{eqnarray}%
The corresponding expressions for the other components are found from the
relations (no summation over $\mu $)%
\begin{eqnarray}
\langle T_{\mu }^{\mu }\rangle _{\mathrm{s}} &\approx &\frac{2\nu }{D-1}%
\langle T_{0}^{0}\rangle _{\mathrm{s}},\;\mu =1,2,\ldots ,D-1,\;  \notag \\
\langle T_{1}^{0}\rangle _{\mathrm{s}} &\approx &\frac{\eta }{r}\frac{%
D-1-2\nu }{D-1}\langle T_{0}^{0}\rangle _{\mathrm{s}}.  \label{T01large}
\end{eqnarray}%
In this limit the diagonal vacuum stresses are isotropic. Note that for a
conformally coupled massless scalar field these leading terms vanish and it
is necessary to keep the next terms in the asymptotic expansion. For
imaginary values of the parameter $\nu $, in the way similar to that used
above for the field squared, it can be seen that the string induced parts in
the diagonal components of the energy-momentum tensor behave as $(\eta
/r)^{D-1}\sin [2|\nu |\ln (\eta /2r)+\psi _{1}]$, where the phase is
different for separate components. For the off-diagonal component the
amplitude of the oscillations decays as $(\eta /r)^{D}$.

\section{Conclusion}

\label{sec5}

In this paper we have investigated quantum effects associated with massive
scalar fields in higher-dimensional dS space in presence of an idealized
cosmic string. In fact we were more interested to calculate the
contributions induced by the conical structure of the sub-space produced by
the string. In order to develop this analysis we have presented the complete
Wightman function as the sum of two contributions: $i)$ The first one
corresponds to the Wightman function on the bulk in the absence of the
string and $ii)$ the second one is induced by the presence of the string.
First in this analysis, we have considered the case where the parameter $q$,
the fractional part of $2\pi $ by the planar angle on the conical surface is
an integer number. In this case the Wightman function is presented as image
sum of the dS Wightman functions, and the VEV for the field squared and the
energy-momentum tensor induced by the string, have been calculated by using
the subtracted Wightman function given in (\ref{Gsubk}) with (\ref{gz}). For
points away from the string the corresponding Wightman function is finite at
the coincidence limit and can be used directly for these evaluations.

For general value of the parameter $q$, we have used the Abel-Plana formula
for the summation over the azimuthal quantum number to provide an explicit
representation for the subtracted Wightman function, Eq. (\ref{WFsub}). Also
for this case we have provided the VEV for the field squared and
energy-momentum tensor in integral forms, formulae (\ref{phi2gen}), (\ref%
{Tmu}), (\ref{T01}). These VEVs depend on the radial and time coordinates in
the combination of $r/\eta $ which is the proper distance from the string
measured in the units of the dS curvature radius $\alpha $. These formulae
are further simplified for points near the string and at large distance from
it. At small distance, the leading order term of the string induced parts in
the VEVs coincide with the corresponding quantities for a massless
conformally coupled field. In this limit the VEVs behave like $(\alpha
r/\eta )^{2-D}$ in the case of $\langle \Phi ^{2}\rangle _{\mathrm{s}}$ and
like $(\alpha r/\eta )^{-D}$ for diagonal components of the energy-momentum
tensor. For the off-diagonal component the leading term vanishes and one has
$\langle T_{1}^{0}\rangle _{\mathrm{s}}\sim (\alpha r/\eta )^{1-D}$. As the
pure dS parts in the VEVs are constants, near the string, the string induced
parts dominate. At large distances from the string and for real values of
the parameter $\nu $, the string induced parts in the VEVs of the field
squared and in the diagonal components of the energy-momentum tensor
monotonically decay as $(\eta /r)^{D-1-2\nu }$. At large distances and for
imaginary $\nu $ the corresponding string induced parts oscillate with the
amplitude going to zero as $(\eta /r)^{D-1}$.

The analysis of VEV associated with scalar and fermionic quantum fields in
the presence of composite topological defects have been presented in \cite%
{Saha1} and \cite{Saha2}, respectively. In these papers a global monopole
and cosmic string have been considered in a higher-dimensional spacetime.
There, by using similar procedure as presented here, we calculated the
contribution induced by the cosmic string on the corresponding vacuum
averages.

\section*{Acknowledgments}

E.R.B.M. thanks Conselho Nacional de Desenvolvimento Cient\'{\i}fico e Tecnol%
\'{o}gico (CNPq) for partial financial support, FAPESQ-PB/CNPq (PRONEX) and
FAPES-ES/CNPq (PRONEX). A.A.S. was supported by the Armenian Ministry of
Education and Science Grant No. 119. A.A.S. gratefully acknowledges the
hospitality of the Abdus Salam International Centre for Theoretical Physics
(Trieste, Italy) where part of this work was done.

\appendix

\section{Integral representation for the Wightman function}

\label{sec:AppWF}

In order to transform the Wightman function, we integrate over the angular
part of the vector $\mathbf{k}$ by using the formula%
\begin{equation}
\int d\mathbf{k}\,\,e^{i\mathbf{k}\cdot \Delta \mathbf{z}}F(k)=(2\pi
)^{(D-3)/2}\int_{0}^{\infty }dk\,k^{D-4}F(k)\frac{J_{(D-5)/2}(k|\Delta
\mathbf{z}|)}{(k|\Delta \mathbf{z}|)^{(D-5)/2}}.  \label{intform}
\end{equation}%
Then we present the product of the Hankel functions in terms of the
MacDonald function,%
\begin{equation}
e^{i(\nu -\nu ^{\ast })\pi /2}H_{\nu }^{(1)}(\lambda \eta )[H_{\nu
}^{(1)}(\lambda \eta ^{\prime })]^{\ast }=\frac{4}{\pi ^{2}}K_{\nu
}(-i\lambda \eta )K_{\nu }(i\lambda \eta ^{\prime }),  \label{HankProd}
\end{equation}%
and use the formula \cite{Wats44}%
\begin{equation}
K_{\nu }(a)K_{\nu }(b)=\frac{1}{2}\int_{0}^{\infty }dx\int_{-\infty
}^{+\infty }dy\,x^{-1}\exp [-2\nu y-abx^{-1}\cosh (2y)-x/2-(a^{2}+b^{2})/2x].
\label{MacProd}
\end{equation}%
As a result the expression for the Wightman function is presented in the
form
\begin{eqnarray}
G(x^{\prime },x) &=&\frac{q\ (\eta \eta ^{\prime })^{(D-1)/2}}{\alpha
^{D-2}(2\pi )^{(D+1)/2}|\Delta \mathbf{z}|^{(D-5)/2}}\int_{0}^{\infty }dp\
p\ \sum_{n=-\infty }^{+\infty }e^{inq\Delta \phi }  \notag \\
&&\times J_{q|n|}(pr)J_{q|n|}(pr^{\prime })\int_{0}^{\infty
}dk\,k^{(D-3)/2}J_{(D-5)/2}(k|\Delta \mathbf{z}|)  \notag \\
&&\times \int_{0}^{\infty }dx\int_{-\infty }^{+\infty }dy\,x^{-1}\exp (-2\nu
y-x/2-\lambda ^{2}\beta /2x),  \label{WF4}
\end{eqnarray}%
with the notation%
\begin{equation}
\beta =2\eta \eta ^{\prime }\cosh (2y)-\eta ^{2}-\eta ^{\prime 2}.
\label{beta}
\end{equation}

The integral over $k$ in Eq. (\ref{WF4}) is taken with the help of the
formula%
\begin{equation}
\int_{0}^{\infty }dk\,k^{(D-3)/2}J_{(D-5)/2}(k|\Delta \mathbf{z}|)\exp
(-k^{2}\beta /2x)=|\Delta \mathbf{z}|^{(D-5)/2}\left( \frac{x}{\beta }%
\right) ^{(D-3)/2}\exp \left( -\frac{|\Delta \mathbf{z}|^{2}x}{2\beta }%
\right) .  \label{Intform3}
\end{equation}%
This leads to the following result%
\begin{eqnarray}
G(x^{\prime },x) &=&\frac{q\ (\eta \eta ^{\prime })^{(D-1)/2}}{\alpha
^{D-2}(2\pi )^{(D+1)/2}}\ \sum_{n=-\infty }^{+\infty }e^{inq\Delta \phi
}\int_{0}^{\infty }dp\ pJ_{q|n|}(pr)J_{q|n|}(pr^{\prime })\int_{-\infty
}^{+\infty }dy\,\exp (-2\nu y)  \notag \\
&&\times 2^{(D-3)/2}\int_{0}^{\infty }du\,u^{(D-5)/2}\exp [-(\beta +|\Delta
\mathbf{z}|^{2})u-p^{2}/4u].  \label{WF5}
\end{eqnarray}

After the integration over $p$ by the formula%
\begin{equation}
\int_{0}^{\infty }dp\ pJ_{q|n|}(pr)J_{q|n|}(pr^{\prime
})e^{-p^{2}/4u}=2u\exp [-(r^{2}+r^{\prime 2})u]I_{q|n|}(2rr^{\prime }u),
\label{Intform4}
\end{equation}%
from (\ref{WF5}) we find%
\begin{eqnarray}
G(x^{\prime },x) &=&\frac{q\ (\eta \eta ^{\prime })^{(D-1)/2}}{\alpha
^{D-2}(2\pi )^{(D+1)/2}}2^{(D-1)/2}\ \sum_{n=-\infty }^{+\infty
}e^{inq\Delta \phi }\int_{-\infty }^{+\infty }dy\,\exp (-2\nu y)  \notag \\
&&\times \int_{0}^{\infty }du\,u^{(D-3)/2}\exp [-(\beta +|\Delta \mathbf{z}%
|^{2}+r^{2}+r^{\prime 2})u]I_{q|n|}(2rr^{\prime }u).  \label{WF6}
\end{eqnarray}%
Now we consider the integral over $y$. Introducing a new integration
variable $z=e^{2y}$ one finds%
\begin{equation}
\int_{-\infty }^{+\infty }dy\,\exp (-2\nu y)\exp (-2\eta \eta ^{\prime
}\cosh (2y)u)=K_{\nu }(2\eta \eta ^{\prime }u).  \label{intform5}
\end{equation}%
Hence, for the Wightman function we find the final expression (\ref{WF}).


\begin{thebibliography}{99}
\bibitem{BD} N. D. Birrell and P. C. W. Davies. \textit{Quantum Fields in
Curved Space} (Cambridge University Press, Cambridge, 1982).

\bibitem{Linde} A. D. Linde. \textit{Particle Physics and Inflationary
Cosmology} (Harwood Academic Publishers, Chur, Switzerland 1990).

\bibitem{Ries} A.G. Riess et al., Astrophys. J. \textbf{659}, 98 (2007);
D.N. Spergel \textit{et al}., Astrophys. J. Suppl. Ser. \textbf{170}, 377
(2007); U. Seljak, A. Slosar, and P. McDonald, J. Cosmol. Astropart. Phys.
\textbf{10}, 014 (2006); E. Komatsu \textit{et al}., arXiv:0803.0547.

\bibitem{Kibble} T. W. Kibble, J. Phys. A \textbf{9}, 1387 (1976).

\bibitem{V-S} A. Vilenkin and E.P.S. Shellard, \textit{Cosmic Strings and
Other Topological Defects} (Cambridge University Press, Cambridge, England,
1994).

\bibitem{Berezinski} V. Berezinski, B. Hnatyk, and A. Vilenkin, Phys. Rev. D
\textbf{64}, 043004 (2001).

\bibitem{Damour} T. Damour and A. Vilenkin, Phys. Rev. Lett. \textbf{85},
3761 (2000).

\bibitem{Bhattacharjee} P. Bhattacharjee and G. Sigl, Phys. Rep. \textbf{327}%
, 109 (2000).

\bibitem{Sarangi} S. Sarangi and S.-H. Henry Tye, Phys. Lett. B \textbf{536}%
, 185 (2002).

\bibitem{Copeland} E. J. Copeland, R. C. Myers, and J. Polchinski, J. High
Energy Phys. \textbf{06}, 013 (2004).

\bibitem{Dvali} G. Dvali and A. Vilenkin, J. Cosmol. Astropart. Phys.
\textbf{03}, 010 (2004).

\bibitem{Cohen} A. G. Cohen and D. B. Kaplan, Phys. Lett. B \textbf{470}, 52
(1999).

\bibitem{Ruth} R. Gregory, Phys. Rev. Lett. \textbf{84}, 2564 (2000).

\bibitem{Cher68} N. A. Chernikov and E.A. Tagirov, Ann. Inst. Henri Poincar%
\'{e} \textbf{9}, 109 (1968); E.A. Tagirov, Ann. Phys. \textbf{76}, 561
(1973).

\bibitem{Cand75} P. Candelas and D. J. Raine, Phys. Rev. D \textbf{12}, 965
(1975).

\bibitem{Dowk76} J. S. Dowker and R. Critchley, Phys. Rev. D \textbf{13},
224 (1976); J. S. Dowker and R. Critchley, Phys. Rev. D \textbf{13}, 3224
(1976).

\bibitem{Bunc78} T. S. Bunch and P. C. W. Davies, Proc. R. Soc. London A
\textbf{360}, 117 (1978).

\bibitem{Birr79} N. D. Birrell, J. Phys. A \textbf{12}, 337 (1979).

\bibitem{Mama81} S. G. Mamayev, Sov. Phys. J. \textbf{24}, 63 (1981).

\bibitem{Vile82} A. Vilenkin and L. H. Ford, Phys. Rev. D \textbf{26}, 1231
(1982).

\bibitem{Alle83} B. Allen, Nucl. Phys. B \textbf{226}, 228 (1983); Ann.
Phys. \textbf{161}, 152 (1985).

\bibitem{Ford85} L. H. Ford, Phys. Rev. D \textbf{31}, 710 (1985).

\bibitem{Kirs93} K. Kirsten and J. Garriga, Phys. Rev. D \textbf{48},567
(1993).

\bibitem{Espo94} G. Esposito, G. Miele, and L. Rosa, Class. Quantum Grav.
\textbf{11}, 2031 (1994).

\bibitem{Inag97} T. Inagaki, T. Muta, and S. D. Odintsov, Progr. Theor.
Phys. Suppl. \textbf{127}, 93 (1997).

\bibitem{Proc03} T. Prokopec and R. P. Woodart, J. High Energy Phys. \textbf{%
10}, 059 (2003); T. Prokopec, O. Tornkvist, and R. P. Woodart, Ann. Phys.
\textbf{303}, 251 (2003); T. Prokopec and R. P. Woodart, Ann. Phys. \textbf{%
312}, 1 (2004); T. Prokopec and E. Puchwein, J. Cosmol. Astropart. Phys.
\textbf{04}, 007 (2004).

\bibitem{Cogn05} G. Cognola, E. Elizalde, S. Nojiri, S. D. Odintsov, and S.
Zerbini, J. Cosmol. Astropart. Phys. \textbf{02}, 010 (2005).

\bibitem{Fine05} F. Finelli, G. Marozzi, G.P. Vacca, and G. Venturi, Phys.
Rev. D \textbf{71}, 023522 (2005).

\bibitem{Dolg06} A. Dolgov and D. N. Pelliccia, Nucl. Phys. B \textbf{734},
208 (2006).

\bibitem{Saha-Set} A. A. Saharian and M. R. Setare, Phys. Let B \textbf{659}%
, 367 (2008).

\bibitem{Bell-Saha} S. Belluci and A. A. Saharian, Phys. Rev. D \textbf{77},
124010 (2008).

\bibitem{Mello} A. A. Saharian, Class. Quantum Grav. \textbf{25}, 165012
(2008); E. R. Bezerra de Mello and A. A. Saharian, J. High Energy Phys.
\textbf{12}, 081 (2008).

\bibitem{scalar} B. Linet, Phys. Rev. D, \textbf{35}, 536 (1987).

\bibitem{scalar1} A. G. Smith, in \textit{Symposium on the Formation and
Evolution of Cosmic String}, edited by G. W. Gibbons, S. W. Hawking and T.
Vachaspati (Cambridge University Press, Cambridge, England, 1989).

\bibitem{scalar2} P. C. Davies and V. Sahni, Class. Quantum Grav. \textbf{5}
1 (1987).

\bibitem{scalar3} T. Souradeep and V. Sahni, Phys. Rev. D \textbf{46}, 1616
(1992).

\bibitem{scalar4} M. E. X. Guimar\~aes and B. Linet, Class. Quantum Grav.
\textbf{10}, 1665 (1993).

\bibitem{scalar5} V. B. Bezerra and E. R. Bezerra de Mello, Class. Quantum
Grav. \textbf{11}, 457 (1994); E. R. Bezerra de Mello, Class. Quantum Grav.
\textbf{11}, 1415 (1994).

\bibitem{ferm} V. P. Frolov and E. M. Serebriany, Phys. Rev. D \textbf{15},
3779 (1287).

\bibitem{ferm1} B. Linet, J. Math. Phys. \textbf{36}, 3694 (1995).

\bibitem{ferm2} E. S. Moreira Jnr., Nucl. Phys. B \textbf{451}, 365 (1995).

\bibitem{ferm3} V. B. Bezerra and N. R. Khusnutdinov, Class. Quantum Grav.
\textbf{23}, 3449 (2006).

\bibitem{Spinelly} J. Spinelly and E. R. Bezerra de Mello, J. High Energy
Phys. \textbf{09}, 005 (2008).

\bibitem{Line86} B. Linet, J. Math. Phys. \textbf{27}, 1817 (1986).

\bibitem{Vega93} H.J. Vega and N. G. Sanchez, Phys. Rev. D \textbf{47}, 3394
(1993).

\bibitem{Ghez02} A. M. Ghezelbash and R.B. Mann, Phys. Lett. B \textbf{537},
329 (2002).

\bibitem{Beze03} E. R. Bezerra de Mello, Y. Brihaye, and B. Hartmann, Phys.
Rev. D \textbf{67}, 124008 (2003); Y. Brihaye and B. Hartmann, Phys. Lett. B
\textbf{669}, 119 (2008).

\bibitem{Basu91} R. Basu, A.H. Guth, and A. Vilenkin, Phys. Rev. D \textbf{44%
}, 340 (1991).

\bibitem{Turo88} N. Turok, Phys. Rev. Lett. \textbf{60}, 549 (1988).

\bibitem{Pru} A. P. Prudnikov, Yu.A. Brychkov, and O.I. Marichev, \textit{%
Integrals and Series }(Gordon and Breach, New York, 1986), Vol. 2.

\bibitem{SahaRev} A. A. Saharian, "The Generalized Abel-Plana Formula with
Applications to Bessel Functions and Casimir Effect," Report No.
ICTP/2007/082 (arXiv:0708.1187).

\bibitem{Saha1} E. R. Bezerra de Mello and A. A. Saharian, Phys. Let. B
\textbf{642}, 129 (2006).

\bibitem{Saha2} E. R. Bezerra de Mello and A. A. Saharian, Phys. Rev. D
\textbf{78}, 045021 (2008).

\bibitem{Wats44} G.N. Watson, \textit{A Treatise on the Theory of Bessel
Functions} (Cambridge, Cambridge University Press, 1944).
\end{thebibliography}
\end{document}